\documentclass[review]{elsarticle}

\usepackage{lineno,hyperref}
\modulolinenumbers[1]

\journal{Astroparticle Physics}

\begin{document}

\begin{frontmatter}

\title{EAS age determination from the study of the lateral 
distribution of charged particles near the shower axis with the ARGO-YBJ experiment}



\author[addressNA1,addressNA2]{B.~Bartoli}
 \address[addressNA1]{Dipartimento di Fisica dell'Universit\`a di Napoli ``Federico II'' \\- Complesso Universitario di Monte Sant'Angelo - via Cinthia - 80126 Napoli - Italy}
 \address[addressNA2]{Istituto Nazionale di Fisica Nucleare - Sezione di Napoli \\- Complesso Universitario di Monte Sant'Angelo - via Cinthia - 80126 Napoli - Italy}

\author[addressLE1,addressLE2]{P.~Bernardini}
 \address[addressLE1]{Dipartimento Matematica e Fisica "Ennio De Giorgi" - Universit\`a del Salento \\- via per Arnesano - 73100 Lecce - Italy}
 \address[addressLE2]{Istituto Nazionale di Fisica Nucleare - Sezione di Lecce \\- via per Arnesano - 73100 Lecce - Italy}

\author[addressIHEP]{X.J.~Bi}
 \address[addressIHEP]{Key Laboratory of Particle Astrophysics - Institute of High Energy Physics - Chinese Academy of Sciences \\- P.O. Box 918 - 100049 Beijing - P.R. China}

\author[addressIHEP]{Z.~Cao}

\author[addressNA1,addressNA2]{S.~Catalanotti}

\author[addressIHEP]{S.Z.~Chen}

\author[addressTIB]{T.L.~Chen}
 \address[addressTIB]{Tibet University - 850000 Lhasa - Xizang - P.R. China}

\author[addressHEB]{S.W.~Cui}
 \address[addressHEB]{Hebei Normal University - Shijiazhuang 050016 - Hebei - P.R. China}

\author[addressYUN]{B.Z.~Dai}
 \address[addressYUN]{Yunnan University - 2 North Cuihu Rd. - 650091 Kunming - Yunnan - P.R. China}

\author[addressLE1,addressLE2]{A.~D'Amone}

\author[addressTIB]{Danzengluobu}

\author[addressLE1,addressLE2]{I.~De Mitri}

\author[addressNA1,addressNA2]{B.~D'Ettorre Piazzoli}

\author[addressNA1,addressNA2]{T.~Di Girolamo}

\author[addressRM2_2]{G.~Di Sciascio}
 \address[addressRM2_2]{Istituto Nazionale di Fisica Nucleare - Sezione di Roma Tor Vergata \\- via della Ricerca Scientifica 1 - 00133 Roma - Italy}

\author[addressShU]{C.F.~Feng}
 \address[addressShU]{Shandong University - 250100 Jinan - Shandong - P.R. China}

\author[addressIHEP]{Zhaoyang Feng}

\author[addressSJU]{Zhenyong Feng}
 \address[addressSJU]{Southwest Jiaotong University - 610031 Chengdu - Sichuan - P.R. China}

\author[addressIHEP]{Q.B.~Gou}

\author[addressIHEP]{Y.Q.~Guo}

\author[addressIHEP]{H.H.~He}

\author[addressTIB]{Haibing Hu}

\author[addressIHEP]{Hongbo Hu}

\author[addressNA1,addressNA2]{M.~Iacovacci}

\author[addressRM2_1,addressRM2_2]{R.~Iuppa}
 \address[addressRM2_1]{Dipartimento di Fisica dell'Universit\`a di Roma ``Tor Vergata'' \\- via della Ricerca Scientifica 1 - 00133 Roma - Italy}

\author[addressSJU]{H.Y.~Jia}

\author[addressTIB]{Labaciren}

\author[addressTIB]{H.J.~Li}

\author[addressIHEP]{C.~Liu}

\author[addressYUN]{J.~Liu}

\author[addressTIB]{M.Y.~Liu}

\author[addressIHEP]{H.~Lu}

\author[addressIHEP]{L.L.~Ma}

\author[addressIHEP]{X.H.~Ma}

\author[addressLE1,addressLE2]{G.~Mancarella}

\author[addressRM3_1,addressRM3_2]{S.M.~Mari}
 \address[addressRM3_1]{Dipartimento di Fisica dell'Universit\`a ``Roma Tre'' \\- via della Vasca Navale 84 - 00146 Roma - Italy} 
 \address[addressRM3_2]{Istituto Nazionale di Fisica Nucleare - Sezione di Roma Tre \\- via della Vasca Navale 84 - 00146 Roma - Italy}

\author[addressLE1,addressLE2]{G.~Marsella}

\author[addressNA2]{S.~Mastroianni}

\author[addressRM2_2]{P.~Montini}

\author[addressTIB]{C.C.~Ning}

\author[addressLE1,addressLE2]{L.~Perrone}

\author[addressRM3_1,addressRM3_2]{P.~Pistilli}

\author[addressPV2]{P.~Salvini}
 \address[addressPV2]{Istituto Nazionale di Fisica Nucleare - Sezione di Pavia - via Bassi 6 - 27100 Pavia - Italy}

\author[addressRM2_1,addressRM2_2]{R.~Santonico}

\author[addressIHEP]{P.R.~Shen}

\author[addressIHEP]{X.D.~Sheng}

\author[addressIHEP]{F.~Shi}

\author[addressLE2]{A.~Surdo\corref{mycorrespondingauthor}}
\cortext[mycorrespondingauthor]{Corresponding author}
\ead{antonio.surdo@le.infn.it}

\author[addressIHEP]{Y.H.~Tan}

\author[addressTO1,addressTO2]{P.~Vallania}
 \address[addressTO1]{Osservatorio Astrofisico di Torino dell'Istituto Nazionale di Astrofisica \\- via P. Giuria 1 - 10125 Torino - Italy}
 \address[addressTO2]{Istituto Nazionale di Fisica Nucleare - Sezione di Torino \\- via P. Giuria 1 - 10125 Torino - Italy}

\author[addressTO1,addressTO2]{S.~Vernetto}

\author[addressTO3,addressTO2]{C.~Vigorito}
 \address[addressTO3]{Dipartimento di Fisica dell'Universit\`a di Torino - via P. Giuria 1 - 10125 Torino - Italy}

\author[addressIHEP]{H.~Wang}

\author[addressIHEP]{C.Y.~Wu}

\author[addressIHEP]{H.R.~Wu}

\author[addressShU]{L.~Xue}

\author[addressYUN]{Q.Y.~Yang}

\author[addressYUN]{X.C.~Yang}

\author[addressIHEP]{Z.G.~Yao}

\author[addressTIB]{A.F.~Yuan}

\author[addressIHEP]{M.~Zha}

\author[addressIHEP]{H.M.~Zhang}

\author[addressYUN]{L.~Zhang}

\author[addressShU]{X.Y.~Zhang}

\author[addressIHEP]{Y.~Zhang}

\author[addressIHEP]{J.~Zhao}

\author[addressTIB]{Zhaxiciren}

\author[addressTIB]{Zhaxisangzhu}

\author[addressSJU]{X.X.~Zhou}

\author[addressSJU]{F.R.~Zhu}

\author[addressIHEP]{Q.Q.~Zhu}

\author{\\(The ARGO-YBJ Collaboration)}


\begin{abstract}
 The ARGO-YBJ experiment, a full coverage extensive air shower (EAS) detector located 
at high altitude (4300 m a.s.l.) in Tibet, China, has smoothly taken data, with very high stability,  
since November 2007 to the beginning of 2013. 
The array consisted of a carpet of about 7000\,m$^2$ Resistive Plate Chambers (RPCs) operated in 
streamer mode and equipped with both digital and analog readout, providing the measurement of particle 
densities up to few particles per cm$^2$.
 The unique detector features (full coverage, readout granularity, wide dynamic range, ..) and 
location (very high altitude) allowed a detailed study of the lateral density profile of charged 
particles at ground very close to the shower axis and its description by a proper lateral 
distribution function (LDF).
 In particular, the information collected in the first 10 m from the shower axis
have been shown to provide a very effective tool for the determination of the shower development stage
(''age'') in the energy range 50 TeV - 10 PeV.
The sensitivity of the age parameter to the mass composition of primary Cosmic Rays is also 
discussed.
\end{abstract}




\vskip 0.3cm

\begin{keyword}
 Cosmic Rays\sep Extensive Air Showers\sep Shower age\sep Composition
 \PACS{98.70.Sa, 96.50.sb, 96.50.sd}
\end{keyword}

\end{frontmatter}



\section{Introduction} 
\label{sec:intro} 
  The energy spectrum and composition in the so-called knee region (10$^{14}$-10$^{16}$\,eV) 
could be crucial to understand the origin and acceleration mechanism of the very high energy 
cosmic ray (CR) flux, which are among the main open problems in particle astrophysics.
 The detailed study of several features (longitudinal structure, lateral distribution of charged
component at detection level, size, etc.) of extensive air showers (EAS), produced by these 
particles in our atmosphere and detected by surface apparata using different techniques, is the 
primary tool for obtaining information about CR primaries when direct measurements 
are prevented by the too low fluxes.
 
 The shower development stage in the atmosphere, 
is expressed by the so-called {\it longitudinal age}, which essentially reflects
the height of the shower maximum with respect to the observation level.
 When measured at a fixed altitude (the detection one), it depends on the energy of the interacting primary particle, 
while, for fixed energy, it depends on the primary nature.
 Indeed, heavier primaries interact
higher in the atmosphere and moreover, according to the superposition principle, they act like a number (equal to A, the
nucleus mass number) of nucleons, each one generating and independent shower at the same height on average (since each
nucleon carries out the same fraction of the nucleus tottal energy).
 The result is that heavy primary nuclei produce showers that, on average, reach their maximum size 
at a greater distance from the detector than a lighter primary of the same energy.
 For this reason, the combined use of shower energy and age estimations can ensure the sensitivity 
to the primary nature.

 An EAS array by itself cannot measure directly the shower development stage, through the 
determination of the depth of the shower maximum, $X_{max}$ (as made for instance by
fluorescence detectors). It can only measure the particle density distribution at ground as 
a function of the core distance (described through a lateral density function, LDF) and from the slope of this 
distribution get information on the longitudinal shower development.
 In fact, the detailed study of the lateral particle density profile at ground is 
expected to provide information on the longitudinal profile of the showers in the atmosphere,
that is to estimate their development stage, or {\it age}.
 The relation between the lateral shape of the detected particle distribution and the shower 
age is quickly explained. Showers starting high in the atmosphere show a flat lateral particle 
distribution, mainly due to multiple scattering processes. Such showers, characterised by a large value
of the age parameter, are called {\it old}. On the contrary, {\it young} showers have started deeper in the
atmosphere, thus having their maximum closer to the observation level. This results in a steeper
lateral particle distribution, which corresponds to a smaller value of the age parameter. 
Apart from fluctuations, the height of the shower maximum depends on energy and mass of the 
initiating particle. Therefore, the lateral shape parameter is also sensitive to the mass of the
CR primary.

 Historically, it was shown that such lateral distribution (at least of the e.m. part of the shower), 
as measured by a traditional sampling EAS array at distances of the order of hundred meters from the core, 
can be properly described by a LDF like the Nishimura-Kamata-Greisen (\emph{NKG}) structure function 
\cite{greisen1956, kamata}, with parameters depending on the shower size, the detection altitude and the shower age. 
The age parameter determined in this way is usually referred to as {\it lateral age} 
\cite{bib:capdev2012, bib:isvhecri2012}, since it is obtained from the LDF. 
It, in principle, coincides with the longitudinal age in particular for purely e.m. showers, 
but they can in fact differ, since most showers come 
from hadrons and the two quantities are measured with completely different techniques.
 However, they are expected to be strongly related.
 Moreover, as experimentally observed, the \emph{NKG} function with a single lateral age parameter is
frequently inadequate to properly describe the lateral density distribution of EAS charged particles
at all distances. This implies that such parameter changes with the radial distance and, for this
reason, the concept of {\it local shower age parameter} was introduced \cite{bib:capdev_lap} to 
denote essentially the lateral age at each point.
As a consequence, any use of the lateral age parameter in order to infer the shower
development stage in the atmosphere has necessarily to face that problem.
To this aim, a full MC simulation of both the shower transportation in the atmosphere and the detailed 
detector response is needed.

 In this paper we show how the peculiar features of the ARGO-YBJ detector can be exploited to study the 
distribution of charged particles in the region around the shower axis by describing its lateral profile 
by means of a proper LDF, thus obtaining an estimation of the shower development stage through the local 
age parameter.
 We also demonstrate and discuss the sensitivity of such age parameter to the masses of the shower initiating 
primaries.

\begin{figure*}[t]
  \centering
   \includegraphics[width=.47\textwidth, height=0.25\textheight]{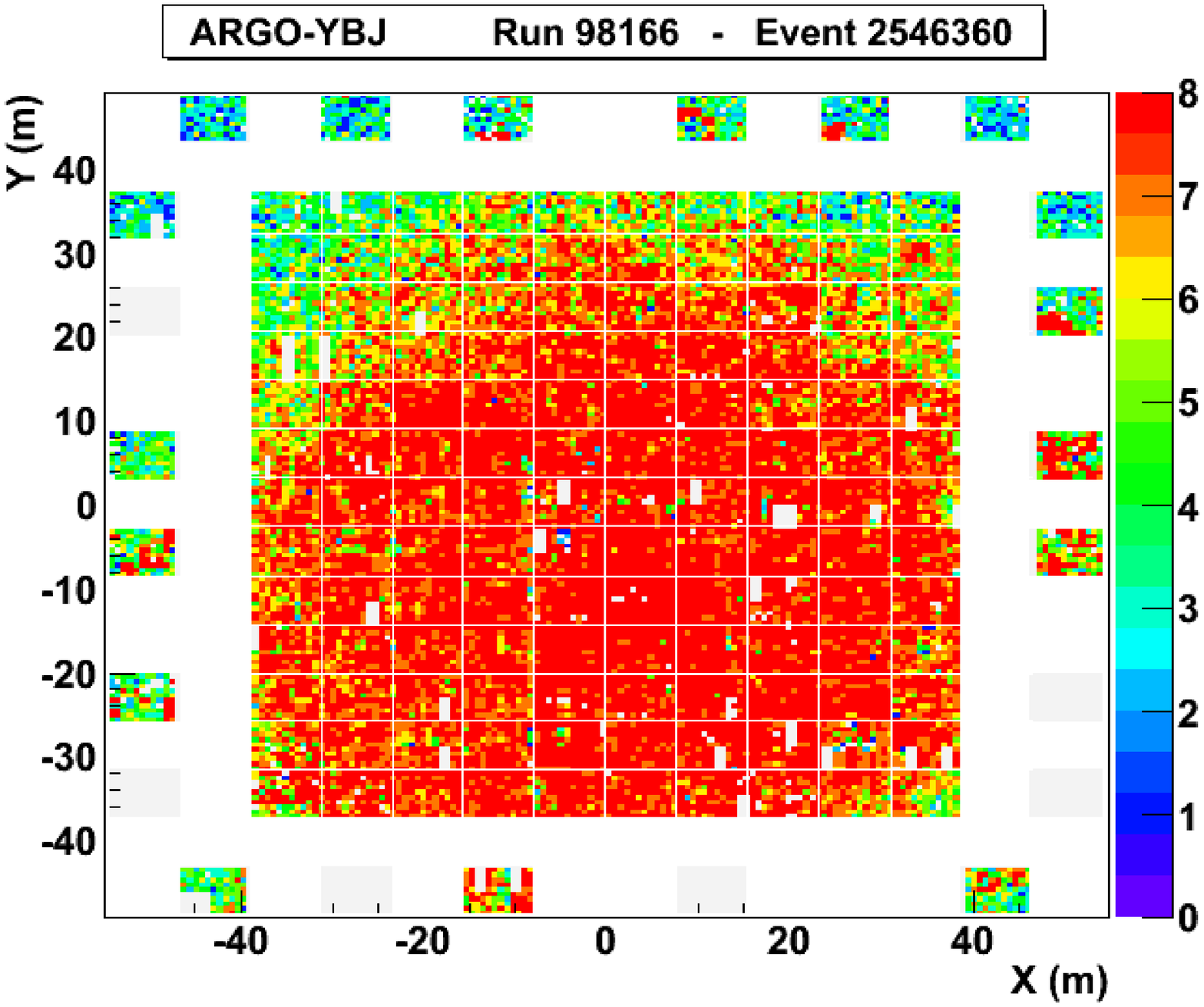}
   \includegraphics[width=.47\textwidth, height=0.25\textheight]{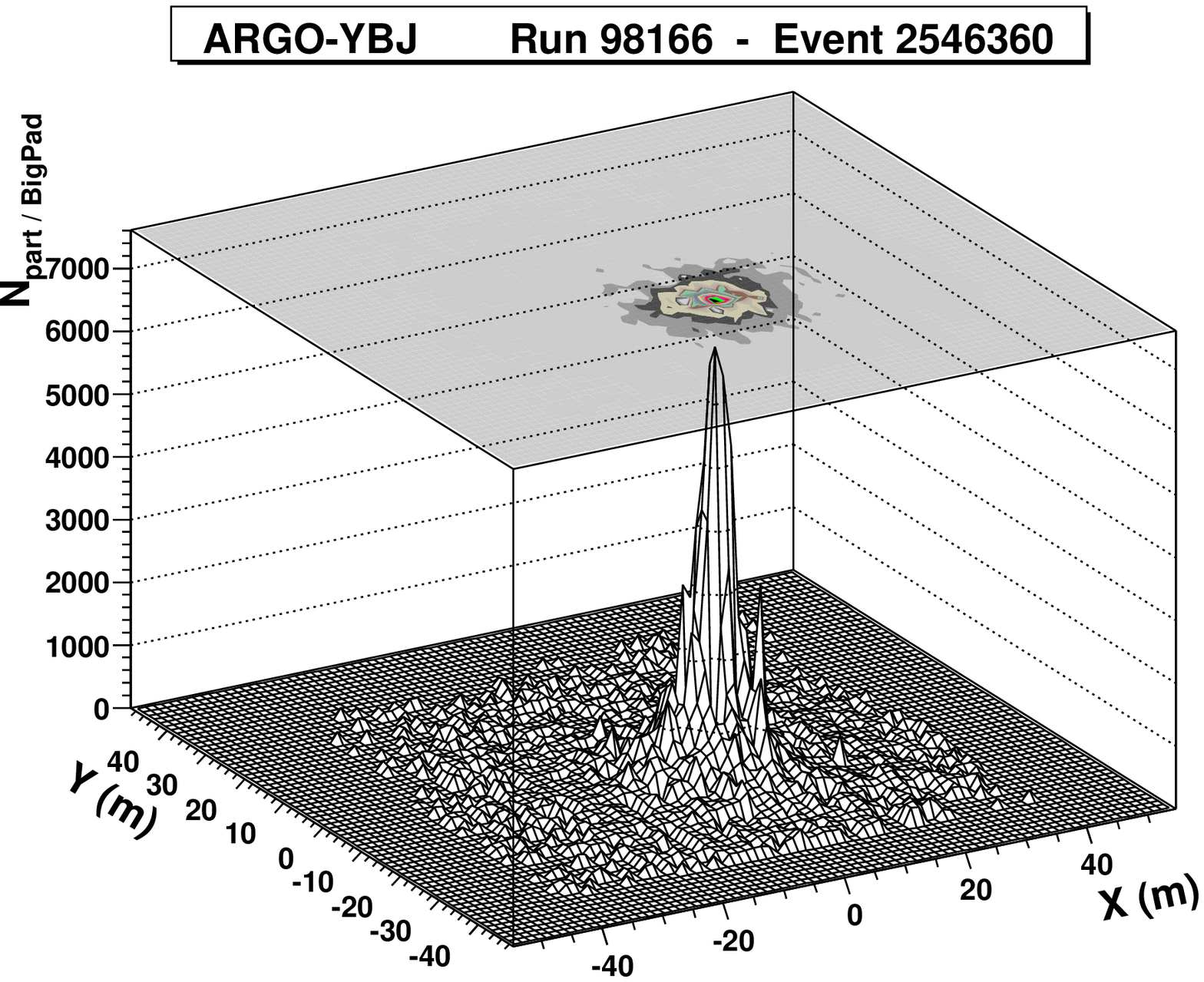} 
\caption[h]{An example of a typical very high energy EAS event recorded by the detector running with the G1 gain scale of 
 the analog system (see text). The hit map at ground is given on the left, the color code representing the strip
multiplicity of each fired pad, evidently saturated on a large portion of the detector. 
The analog RPC signal ($N_{part}/BigPad$) vs the position on the detector is shown on the right:
the core and the lateral particle distribution of the shower in the first few 
meters from it are clearly visible.}
 \label{fig:frontstruc}
\end{figure*}

\section{The ARGO-YBJ experiment}
\label{sec:argo}
%
  The ARGO-YBJ detector was a full coverage extensive air shower (EAS) array made by a single layer of 
Resistive Plate Chambers (RPCs) operated in streamer mode, for \hbox{$\gamma$-astronomy} observations with 
$\sim 100$ GeV energy threshold, search of Gamma Ray Bursts in the full GeV/TeV energy range and 
CR studies in the energy range (1-10$^{4}$)\,TeV \cite{Argo_Results}.  
 For these purposes, the array was installed in the Cosmic Ray Observatory of YanBaJing (Tibet, China), at an altitude of 
4300$\,$m above sea level (corresponding to a vertical atmospheric depth of about $606 \,$g/cm$^2$), and ran in its full 
configuration since November 2007 until February 2013. 
It was organized in 153 clusters of 12 RPCs each. Any single RPC was read out by ten \hbox{$62 \times 56 \,$cm$^2$} 
pads, which were further divided into 8 strips, thus providing a larger particle counting 
dynamic range \cite{bib:bacci2002, bib:aielli2006}.
The signals coming from all the strips of a given pad were sent to the same channel of a multi-hit TDC.  
The whole system provided a single hit (pad) time resolution of $\sim$1.8 ns, which, joined to the full 
coverage feature, allowed a complete and detailed three-dimensional reconstruction of the shower front with 
unprecedented space-time resolution. 
A system for the RPC analog charge readout \cite{bib:iacovICRC2015} from larger pads,
each one covering half a chamber (the so called {\it big pads}, BP), has also been implemented and 
took data since January 2010.
This actually extended the detector sensitivity range from about 10$^{14}\,$eV up to about 10$^{16}\,$eV of
the primary energy. 
The analog readout system has been operated with different gain scales (from G0 to G7, with increasing gains), 
which determined the threshold and the maximum number of particles that could be reliably measured by each BP. 
The highest gain scale G7 allowed low density values to be measured down to few particles per m$^2$, overlapping 
its dynamic range with the detector operated in 'digital mode', i.e. simply counting the number of fired 
strips, that saturates at about 20/m$^2$. The data collected by this scale were mainly used for calibration purposes, 
following the procedure described in \cite{bib:iacovICRC2015}.
The other scales had decreasing gains (down to G0) and allowed measuring ever increasing 
densities up to several $10^4$/m$^2$ \cite{bib:iacovICRC2015}.
In Fig.\,\ref{fig:frontstruc}, the three-dimensional histogram of a EAS event imaged through the analog 
readout is shown and compared to the one obtained by the digital readout; the core is clearly identified and the shower front structure in the
first few meters can be resolved.
These features allowed to study, for the first time, the detailed profile of the particle
density distribution at the observation level even very close to the shower axis.
 
 The possibility of a detailed investigation of the distribution of particles detected in the first few meters from the shower 
axis provides a new and efficient way of selecting events initiated by light mass primaries (see last Section), 
without relying on the muon signal.
 On the other hand, such a study could give new inputs, in the very forward kinematic region, to the hadronic interaction 
models currently used for the investigation of the cosmic ray flux and origin at the highest energies.

\begin{figure}[t]
	\centerline{\includegraphics[width=0.7\textwidth, height=6.5cm]{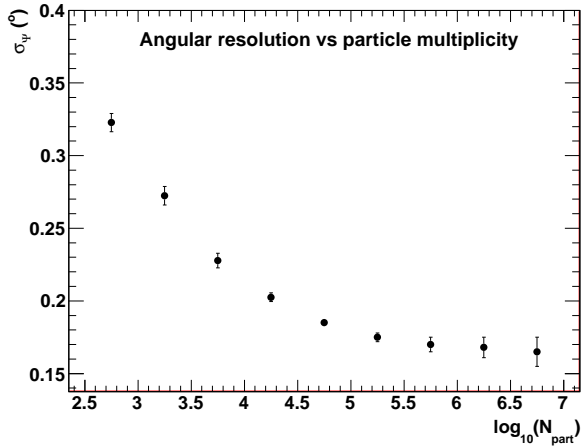} }
 
 \caption{Angular resolution of incoming direction reconstruction for showers triggering the analog system, as a 
          function of $log_{10}(N_{part})$, the logarithm of particle multiplicity on the whole central detector.
	  The angular resolution is here defined as $\psi_{70}/1.58$, where $\psi_{70}$ is the space
	  angle including the 70\% of reconstructed directions with respect to the true one.}
\label{fig:ang_res}
\end{figure}
\begin{figure}[t]
   \centerline{\includegraphics[width=0.7\textwidth, height=6.5cm]{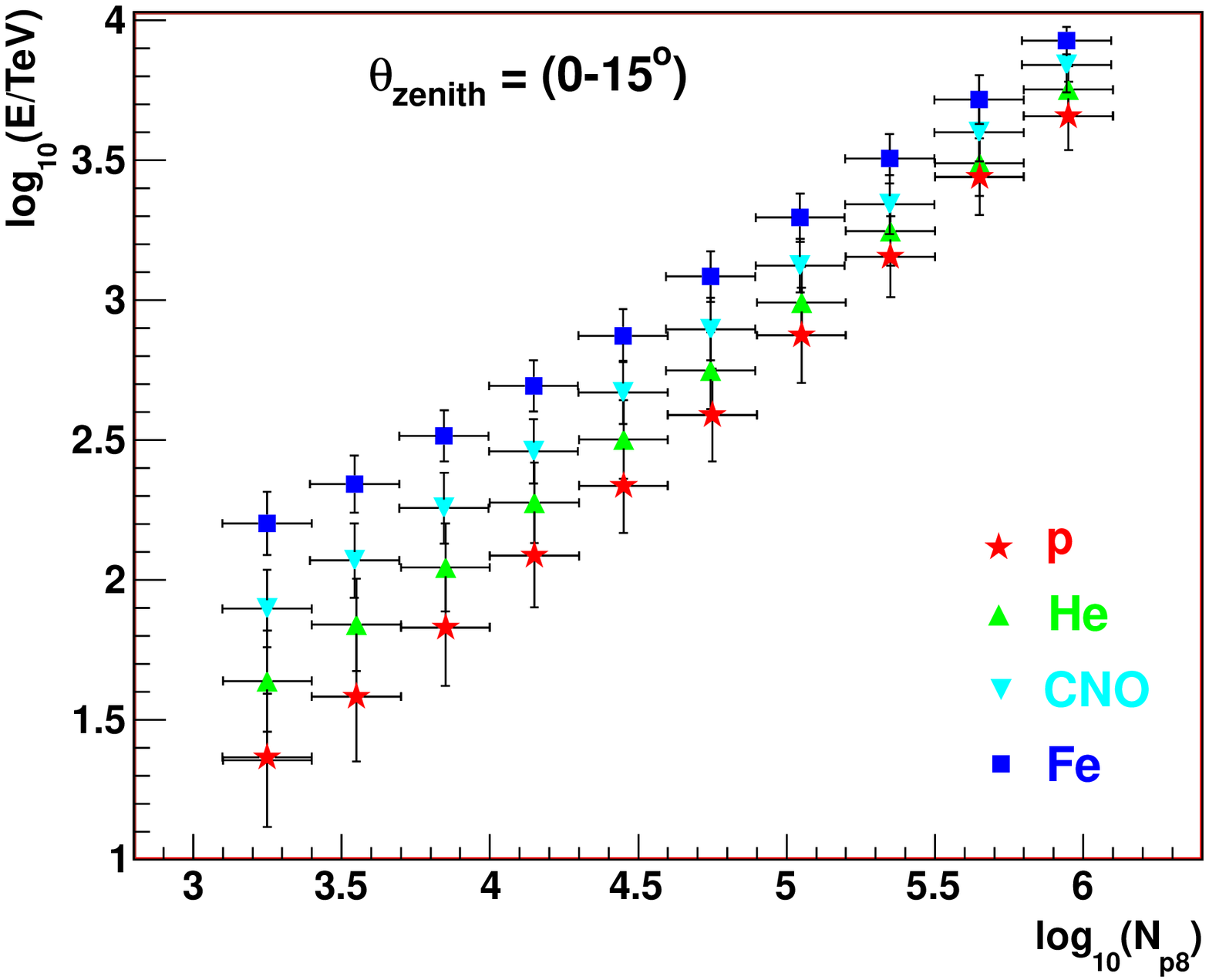} }
   \caption{Average primary energy for p, He, CNO group and Fe initiated MC showers, reconstructed in the zenith angle 
            range $\theta_{zenith}<$ 15$^\circ$, for various intervals of $N_{p8}$,
            the number of detected particles within 8\,m from the shower axis.}
  \label{fig:EvsNp8}
\end{figure} 

\section {Monte Carlo simulations and event selection}
\label{sec:performance}
%
 For the analysis presented here, several air shower samples induced by proton, He, CNO elemental group 
and Fe primaries have been simulated, for a total amount of several millions of events in the (10$^{12}$-10$^{16}$)\,eV 
energy range.
 The simulated showers were produced by using the {\it CORSIKA} code \cite{corsika}, with 
{\it QGSJET-II.03} \cite{qgsjet} as hadronic interaction model, while {\it FLUKA} code \cite{fluka1, fluka2}
has been used at lower energies.
The showers were generated in the zenith angle range $\theta <45^{\circ}$, 
according to a spectral index -1 and subsequently weighted in such a way to follow the flux normalizations and spectra
as given in \cite{hoerandel2003}.
 Throughout the whole paper, if not differently specified, we adopted the H\"orandel model to obtain a mixed sample of the
above cited elements.
 With cores randomly sampled in a larger area (about ten times) than the detector surface, such showers have 
been given in input to a {\it GEANT} \cite{geant3} based program fully simulating the detector 
structure and response (including the effects of time resolution, trigger logic, electronics noise, 
readout system, etc.). The Monte Carlo (MC) events triggering the analog system readout ($\geq$73 fired pads in
a cluster) have then been processed by the same reconstruction program used for real data. 

\begin{figure}[t]
	\centerline{\includegraphics[width=0.7\textwidth, height=6.5cm]{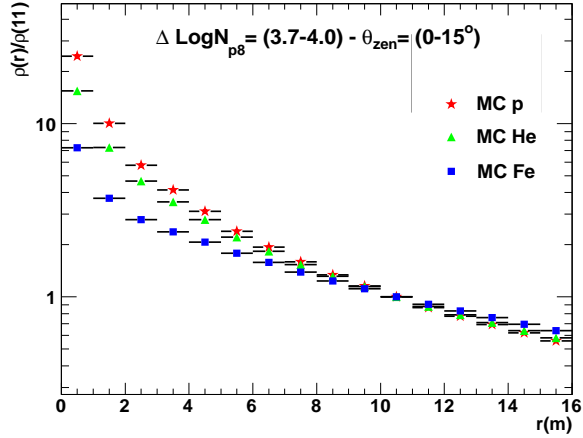} }
 
 \caption{Lateral density distribution of particles around the shower axis for MC proton, helium and iron 
          initiated showers, with $ 10^{3.7}< N_{p8} < 10^{4.0}$ and $\theta_{zenith}<$ 15$^\circ$.
	  The distributions are normalized to the density at 11\,m of distance from the 
	  shower axis.}
\label{fig:Ldf_pHeFe}
\end{figure}

The shower core was determined through the Maximum Likelihood method applied to the lateral density distribution of the 
detected particles, which was fitted to a modified NKG function (see, e.g., \cite{moonshadow2011}). 
 The algorithm ensured an accuracy of the order of 1\,m or less.
Concerning the reconstruction of incoming direction, the front of each detected shower was fitted by 
a conical shape, with vertex in the core position and aperture as a free parameter of the fit. 
The resulting angular resolution 
is found to be much better than 1 degree for all events triggering the analog system, with a gradual
and constant improvement up to the highest hit multiplicities, as shown in Fig.\,\ref{fig:ang_res}.
The events were subsequently selected by requiring the core position
to be in a fiducial area of $64 \times 64$\,m$^{2}$ around the detector center.
This work was also restricted to events with reconstructed zenith angle $\theta_{zenith}<15^{\circ}$.

The study of the MC events (see \cite{hadro_proc}) allowed to identify the truncated size 
$N_{p8}$, defined as the number of particles detected within a distance of 8\,m from the 
shower axis, as a suitable, although mass dependent, estimator of the primary CR energy (Fig.\,\ref{fig:EvsNp8}), since well 
correlated to the total shower energy, not biased by effects due to the finite detector size, nor 
dominated by shower to shower fluctuations. As a consequence, in order to select event samples in given, 
sufficiently narrow, intervals of energy, specific bins of such observable have been extensively used 
all over this analysis.

 Concerning the experimental dataset, the G1 and G4 gain scale samples were used for this analysis, apart the G7 one used 
for calibration pourposes. This led to define a specific cut for the maximum particle number (hereafter $N_{max}$) on a single 
big pad in each event, in the G4 and G1 samples separately, in order to select showers with a lateral particle density 
distribution well above the threshold set by the analog system:
$log_{10}(N_{max})>1.7$ for G4 and 
$log_{10}(N_{max})>2.7$ for G1 scale 
(being $N_{max}$ the maximum particle number measured by a single big pad in the event). 
Moreover, a cut on $N_{p8}$ ensured to avoid possible saturation effects on each of the two gain
scales: $3<log_{10}(N_{p8})<5$ for G4 and $log_{10}(N_{p8})>4$ for G1.
 The same fiducial cuts used for real data were finally applied to the simulated samples.

\begin{figure*}[t]
  \centering
   \includegraphics[width=0.45\textwidth, height=0.2\textheight]{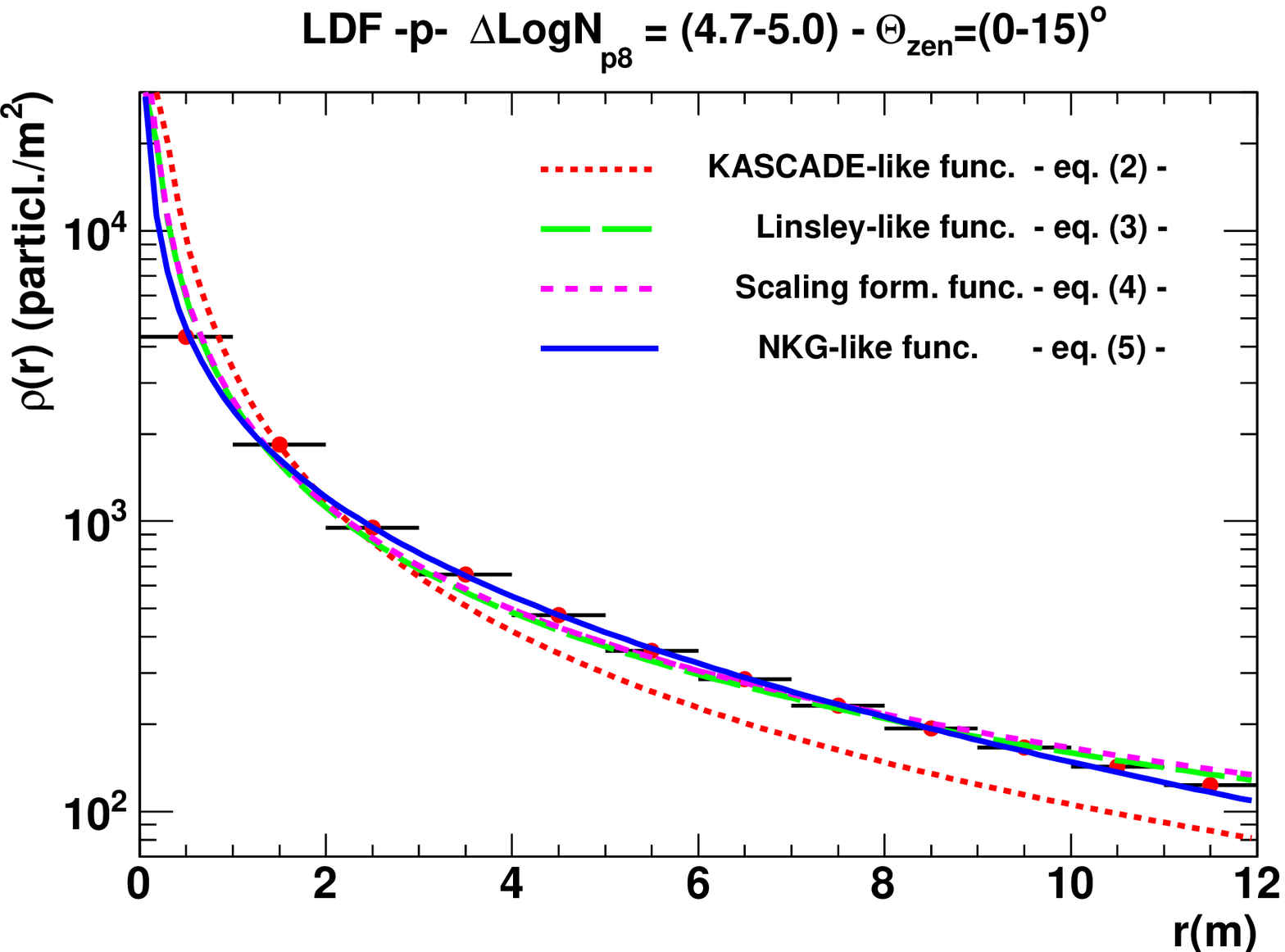}
   \includegraphics[width=0.45\textwidth, height=0.2\textheight]{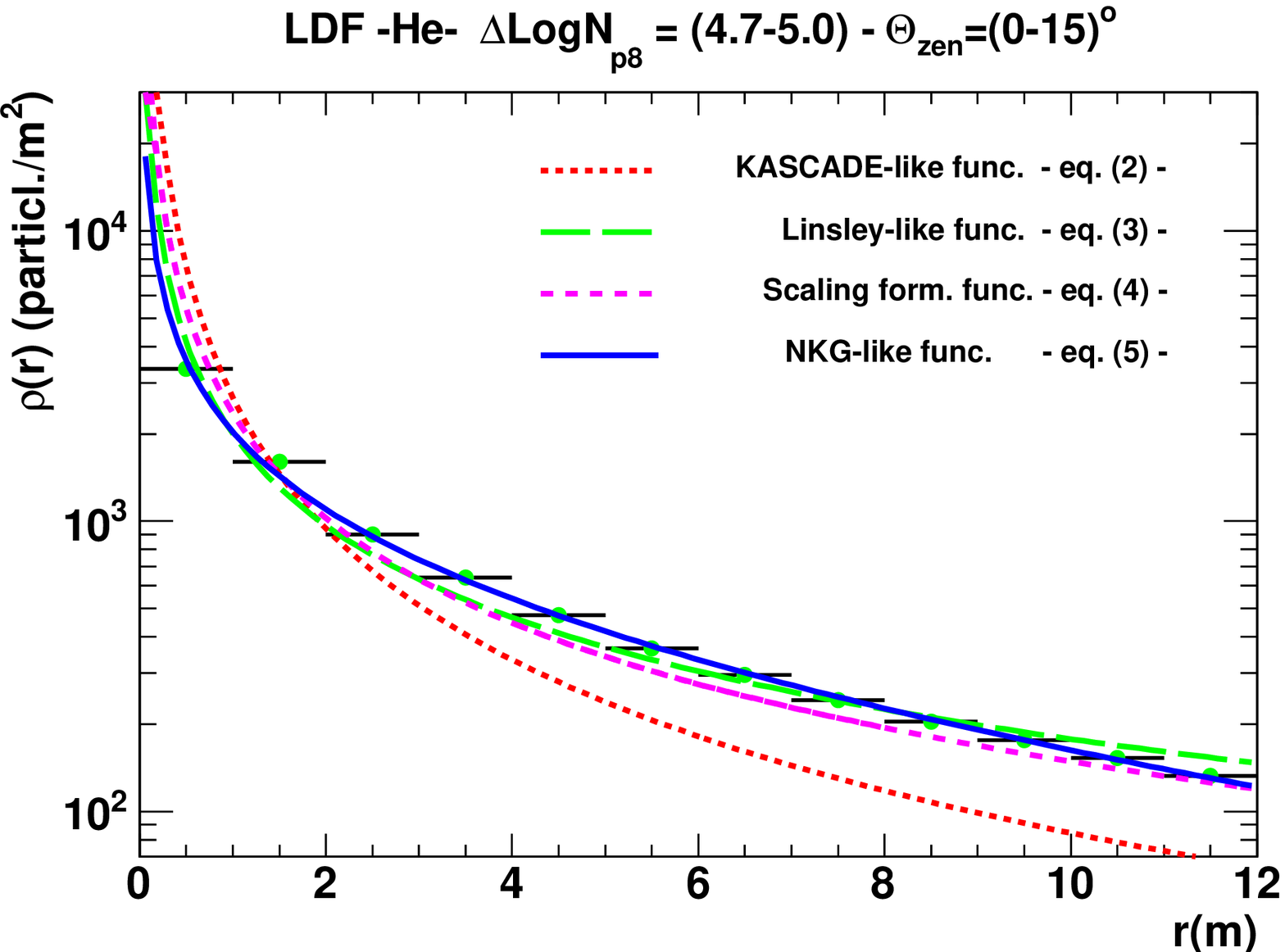}
   \includegraphics[width=0.45\textwidth, height=0.2\textheight]{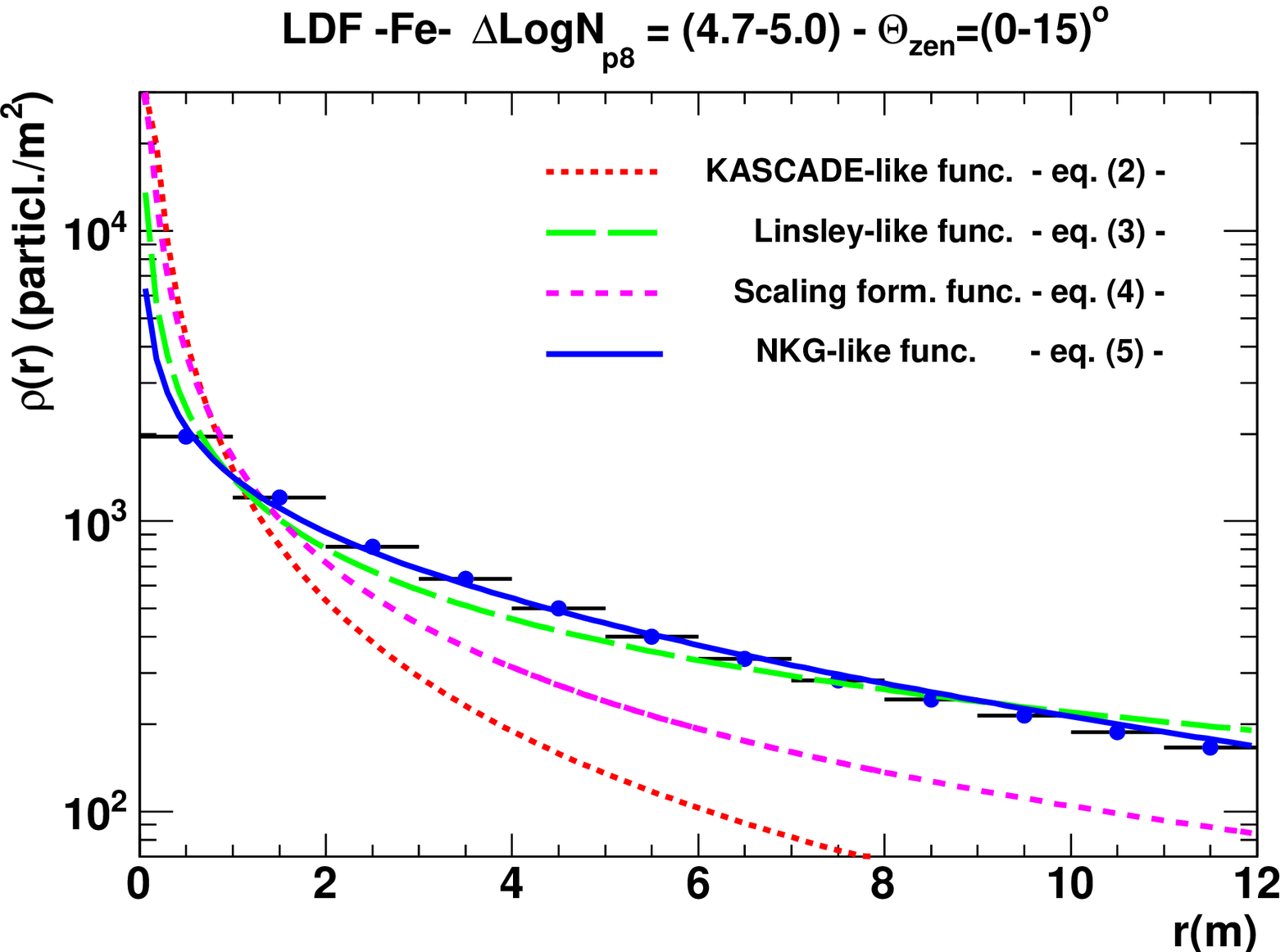}
	  \vskip -0.2cm  
 \caption{Reconstructed average density distribution of the detected particles around the shower axis for simulated 
          proton, helium and iron initiated showers (as indicated in the label on top of each plot), with
	  zenith angle $\theta < 15^{\circ}$ and $10^{4.7} < N_{p8} < 10^{5.0}$
          (corresponding to about $500\,$TeV, $700\,$TeV, and $1.4\,$PeV 
	  for primary p, He, and Fe, respectively \cite{hadro_proc}). 
          The fits with LDFs in Eq.\,\ref{eq:nkg_kasc}, Eq.\,\ref{eq:linsley} with $R_M$ left as free parameter, 
          Eq.\,\ref{eq:lagutin}, and Eq.\,\ref{eq:nkglikeeq} are superimposed (see text).  }
     \label{fig:fit_various_func}
\end{figure*}

\begin{figure*}[t!]
  \centering
   \includegraphics[width=0.45\textwidth, height=0.2\textheight]{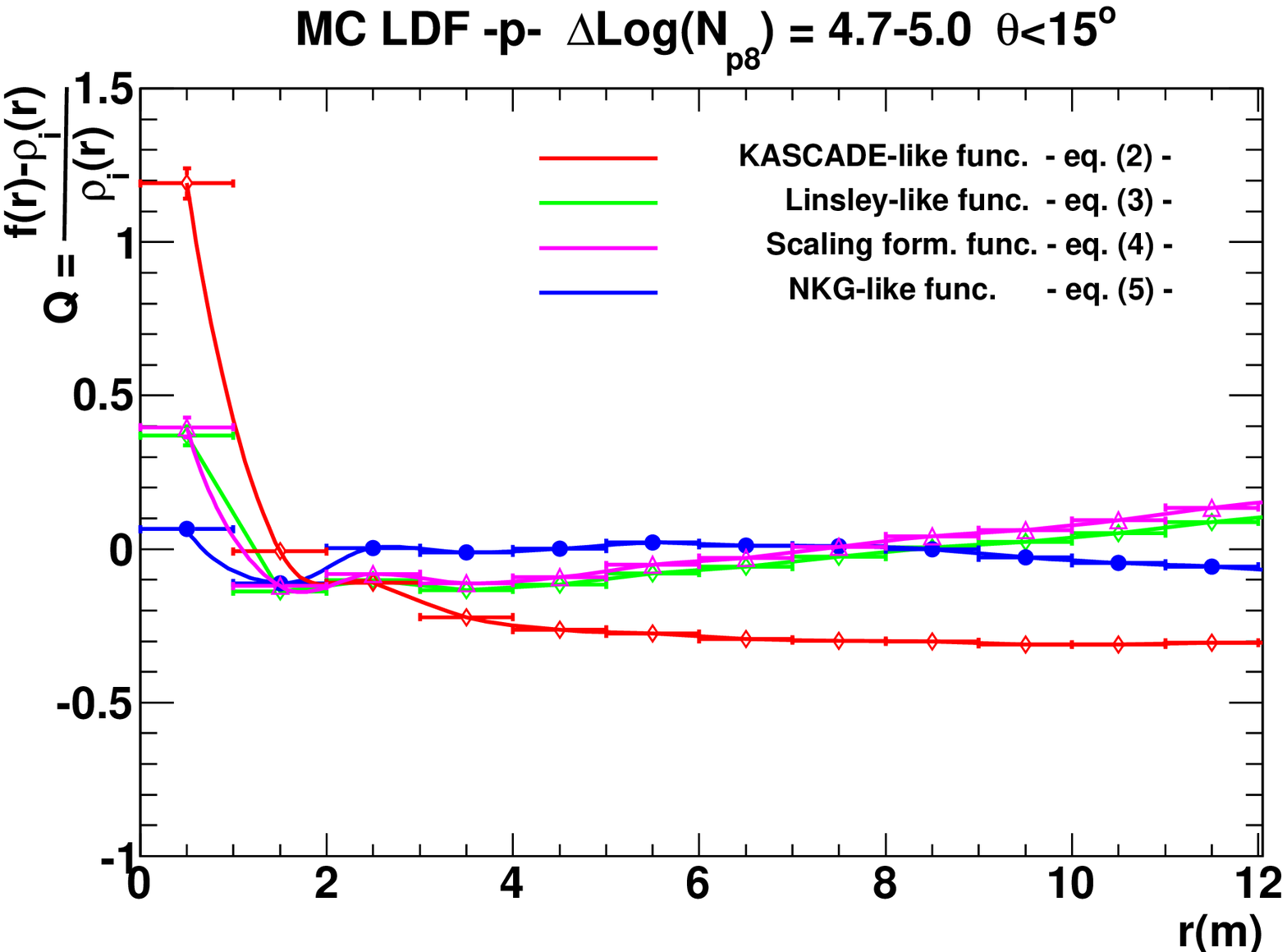}
   \includegraphics[width=0.45\textwidth, height=0.2\textheight]{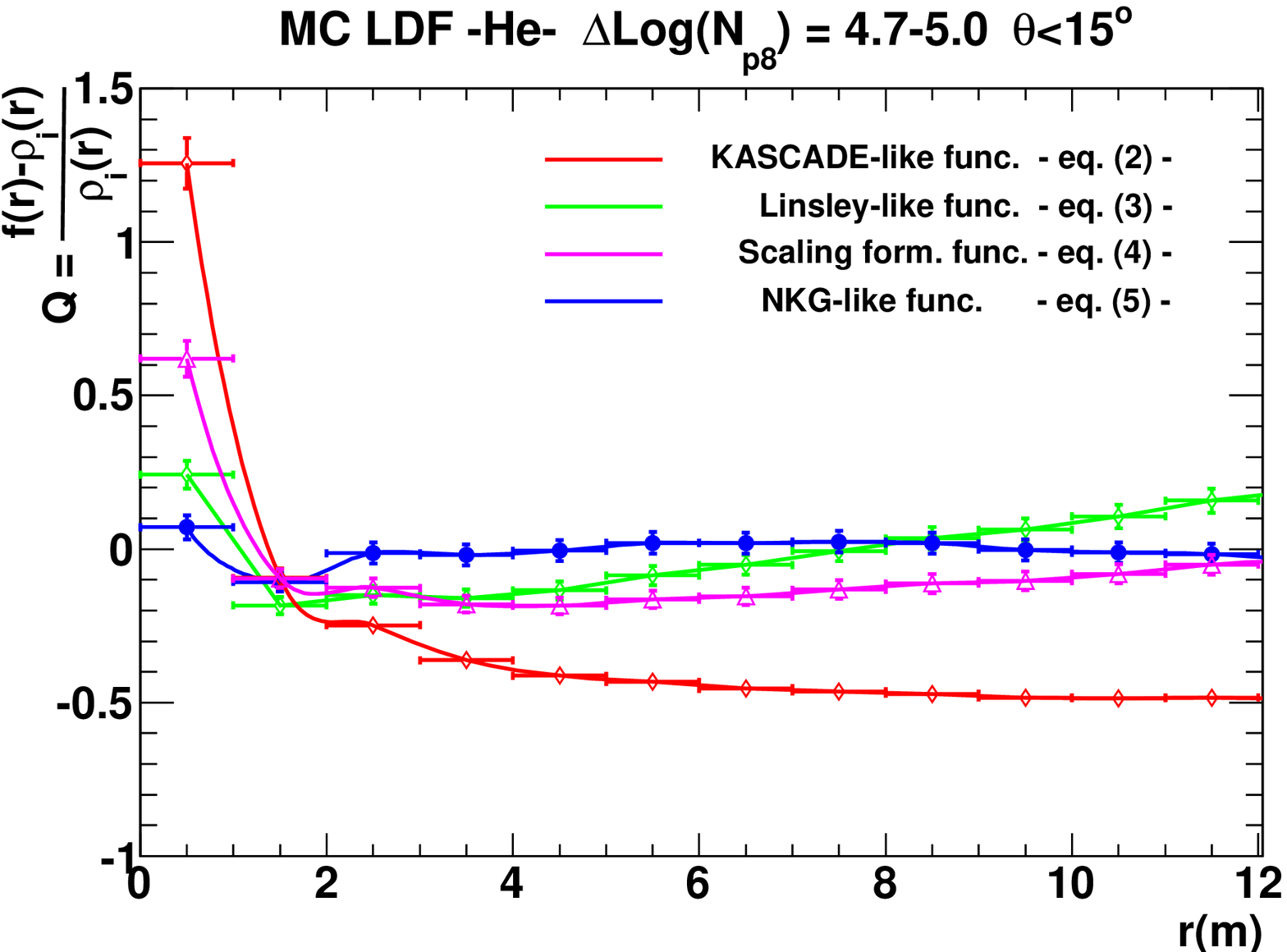}
   \includegraphics[width=0.45\textwidth, height=0.2\textheight]{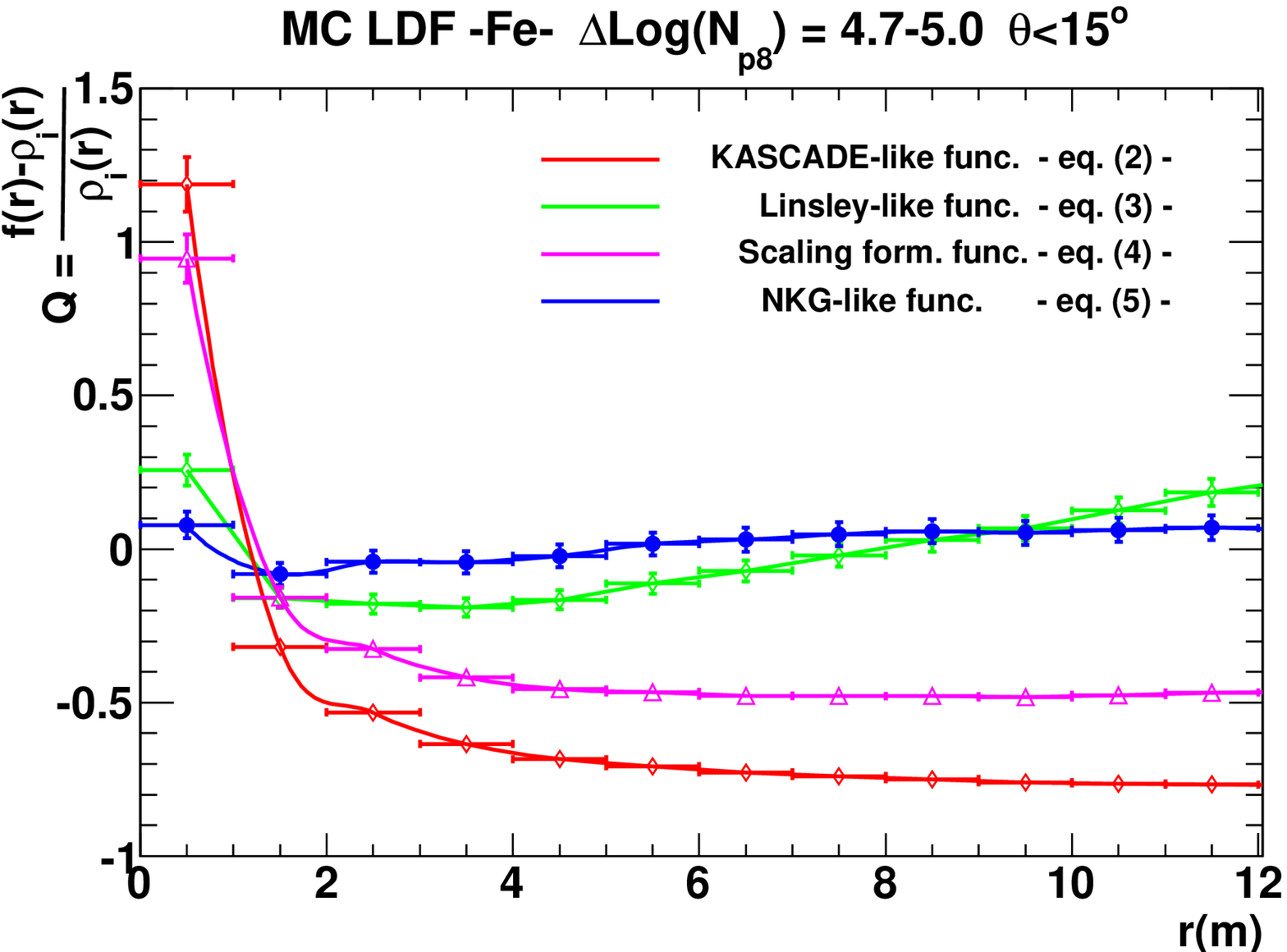}
	  \vskip -0.2cm  
  \caption{Residuals for the fits of the average LDF of simulated proton, helium and iron initiated extensive air 
           showers as shown in Fig.\protect\ref{fig:fit_various_func}.}
  \label{fig:quality_fit_p}
\end{figure*}

\section{Parametrizations of the LDF}
\label{sec:study}
%
As stated in the Section \ref{sec:intro}, the basic idea is to get information on the shower development 
stage from the lateral density distribution structure around the shower axis.
 Thus, the average particle distributions starting from the reconstructed core have been studied in detail 
for each $N_{p8}$ bin and, in the case of simulations, for different primary types.
 As an example, the Fig.\,\ref{fig:Ldf_pHeFe} shows the average lateral distribution of particles obtained 
for a sample of simulated proton induced shower events in the 
interval $\Delta N_{p8}$\,=\,10$^{3.7}-10^{4.0}$, corresponding to an average energy $E_{p}$\,$\simeq$70$\,$TeV,
together with the average LDFs.
 Moreover, the analogous distributions for He and Fe primaries in the same $\Delta N_{p8}$ interval 
(corresponding to an average energy $E_{He}$\,$\simeq$100$\,$TeV and $E_{Fe}$\,$\simeq$300$\,$TeV, 
respectively) are also shown in the same plot. 

The above cited NKG function, which describes fairly well the lateral distribution of charged particles,
especially over distancies of hundreds meters at the 
observation level, has the following mathematical form:

\vskip -0.25cm
\begin{equation}
  \label{eq:nkg}
  \rho_{1}(r)= N_e \ C(s) \Big( \frac{r}{R_{M}}\Big)^{s-2} \Big( 1 + \frac{r}{R_{M}}\Big)^{s-4.5}		  
\end{equation}

\noindent 
Here $\rho_{1}(r)$ is the particle density at a distance $r$ from the shower axis, 
$N_e$ is total number of particles at the observation depth, $C(s)$ is given by

\vskip -0.3cm
$$ C(s)= \frac{1}{2\pi R_{M}^{2}} \times \frac{\Gamma(4.5-s)}{\Gamma(s)\Gamma(4.5-2s)} $$

\noindent 
being $\Gamma(x)$ the gamma function, $R_{M}$ the Moli\`ere radius at ground, 
$s$ the lateral age parameter.

\begin{figure}[t]
  \centerline{\includegraphics[width=0.7\textwidth, height=6.5cm]{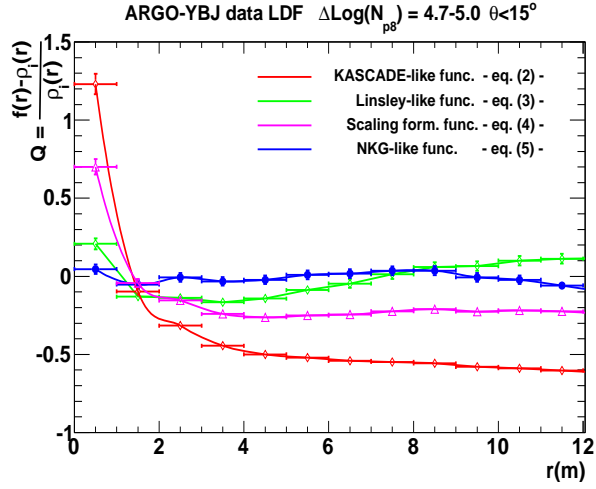} }
   \caption{Residuals from the fits with the LDF functions in Eq.\,\ref{eq:nkg_kasc}, Eq.\,\ref{eq:linsley} with $R_M$ 
            left as free parameter, Eq.\,\ref{eq:lagutin} and Eq.\,\ref{eq:nkglikeeq} to the average lateral density 
	    distribution of particles around the shower axis, for the ARGO-YBJ data (G1 gain scale, see text), 
	    with $10^{4.7} < N_{p8} < 10^{5.0}$ and $\theta < 15^{\circ}$. }
  \label{fig:quality_data}
\end{figure} 

 Several modifications of the \emph{NKG} form were proposed as LDFs in order to better reproduce the data of 
various experiments that measured particle densities at ground.
This could be done by introducing the concept of local age $s=s(r)$ \cite{bib:capdevielle}, 
or suitably modifying the original form 
given in Eq. \ref{eq:nkg}.
As an example, the KASCADE experiment \cite{KASCfunc} used a \emph{NKG-like} function able 
to describe the measured lateral distribution for showers with energies up to $10^{17}$ eV 
and for core distances up to 200\,m: 
\begin{equation} 
 \label{eq:nkg_kasc}
 \rho_{2}(r)= N_e C(s) \Big( \frac{r}{r_0}\Big)^{s-\alpha} \Big( 1 + \frac{r}{r_0}\Big)^{s-\beta}
\end{equation}
\noindent
where
$$ C(s)= \frac{1}{2\pi r_0^{2}} \times 
         \frac{\Gamma(\beta-s)}{\Gamma(s-\alpha+2)\Gamma(\alpha+\beta-2s-2)} $$
\noindent 
In this case some parameters have been optimized with Monte Carlo data, 
with $\alpha=1.5$, $\beta = 3.6$ and $r_0 = 40\,$m being used as radial scale factor.
Another example is given by the AGASA  group \cite{linsleyfuncAGASA} which used a generalized \emph{NKG} 
function with an additional term to take into account density measurement at very large distances, 
inspired by a function suggested by Linsley \cite{linsleyfunc}:
\begin{equation}
  \label{eq:linsley}
    \rho_{3}(r) = C \left( \frac{r}{R_M} \right)^{-1.2} \left( 1+\frac{r}{R_M}\right)^{-(\eta-1.2)} 
                  \left[ 1.0+\left(\frac{r}{1000 m}\right)^2  \right]^{-\delta}
\end{equation}
\noindent
where $C$, $\eta$ and $\delta$ are free parameters.
This function describes well lateral distribution of charged particle up to distances of several 
km from the shower core. 
The $\eta$ parameter is related to the LDF slope and depends on the zenith angle.
A different approach is the so called \emph{scaling formalism} \cite{raikin2}. In this case
\begin{equation}
  \label{eq:lagutin}
  \rho_{4}(r)= \frac{N_e}{r^2_0} C \left(\frac{r}{r_0}\right)^{-\alpha} 
               \left(1+\frac{r}{r_0}\right)^{-(\beta-\alpha)}
                 \left[ 1 + \left( \frac{r}{10r_0}\right)^{2}\right]^{-\delta}
\end{equation}
\noindent 
where $C= 0.28$, $\alpha=1.2$, $\beta=4.53$, $\delta=0.6$, while $r_0$ here becomes a 
free parameter that is shown to be correlated with the shower age. It has to be noticed that this function  
describes well particle densities measured far from the core, like in the AGASA experiment \cite{bib:fomin}. 

\begin{figure}[t!]
  \centering
  \includegraphics[width=0.7\textwidth, height=6.5cm]{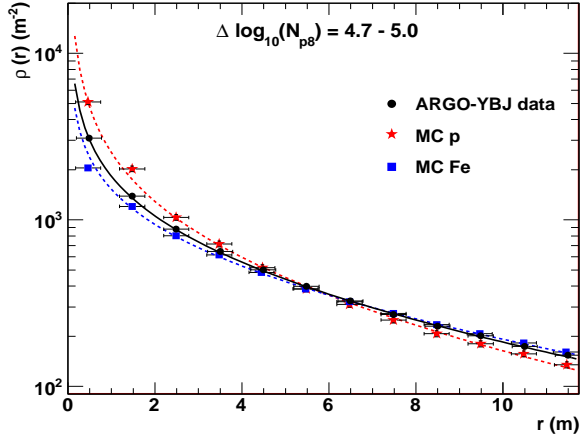}
   \caption{Reconstructed lateral density distribution of the particles around the shower axis for 
            ARGO-YBJ data with $10^{4.7} < N_{p8} < 10^{5.0}$ and $\theta < 15^{\circ}$. 
            The superimposed fit with the \emph{NKG-like} LDF in Eq.\,\ref{eq:nkglikeeq} is also shown 
	    (solid line). 
	    The experimental data distribution lies between the similar MC distributions (shown 
	    for comparison with dashed fit lines) from pure proton and iron induced showers.}
 \label{fig:nkg_data}
\end{figure}

For ARGO-YBJ data, the lateral particle distributions were firstly fitted with each of the different 
parametrizations above reported (Fig.\,\ref{fig:fit_various_func} and \ref{fig:quality_fit_p}).
 Such a systematic study showed that no one of those
functions was able to fit ARGO-YBJ data in a satisfactory way. In particular, the fit with the original 
\emph{NKG} formula (Eq.\ref{eq:nkg}) 
did not give good results unless using $R_M$ values much lower than the actual 
Moli\`ere radius at the experimental site.

\begin{figure}[t]
  \centerline{\includegraphics[width=0.7\textwidth, height=6.5cm]{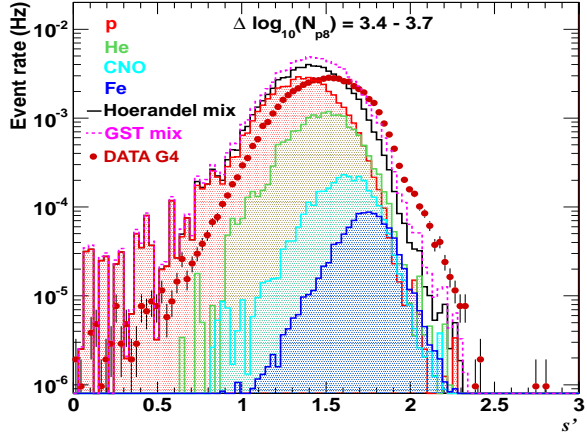} }
 \caption{The $s'$ parameter distribution in the multiplicity bin $10^{3.4} < N_{p8} < 10^{3.7}$.
          The filled areas are MC distributions of protons (red), Helium (green), CNO (cyan) and Fe (blue). 
          Red dots are the $s'$ distribution for ARGO-YBJ dataset (from G4 gain scale) in the same $N_{p8}$ bin. 
          The H\"orandel model \cite{hoerandel2003} (black solid line) is used in weighting individual element 
	  distributions. The GST model \cite{GST2013} mixed distribution (magenta dashed line) is also shown in 
	  the plot for comparison. 
	  (For interpretation of the references to colour in this figure legend, the reader is referred to the 
	  web version of this article.)}
 \label{s_distr_G4}
\end{figure}
\begin{figure}[t]
  \centerline{\includegraphics[width=0.7\textwidth, height=6.5cm]{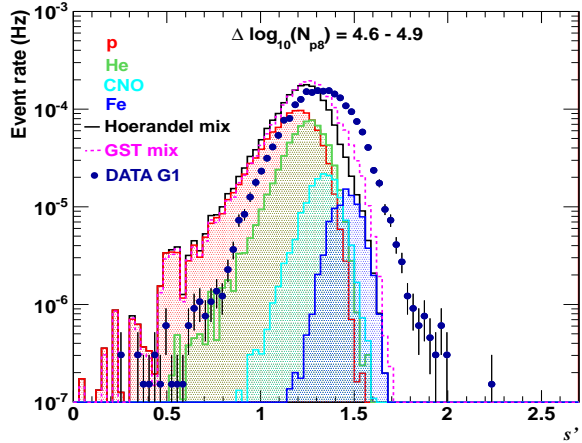} }
 \caption{The $s'$ parameter distribution in the multiplicity bin $10^{4.6} < N_{p8} < 10^{4.9}$.
          The filled areas are MC distributions of protons (red), Helium (green), CNO (cyan) and Fe (blue). 
          Dark-blue dots are the $s'$ distribution for ARGO-YBJ G1-dataset in the same $N_{p8}$ bin. 
	  The H\"orandel model (black solid line) is used in weighting individual element distributions.
	  The GST model mixed distribution (magenta dashed line) is also shown in the plot for comparison.
	  (For interpretation of the references to colour in this figure legend, the reader is referred to the 
	  web version of this article.)}
 \label{s_distr_G1}
\end{figure}

 The best performance in describing both simulated and experimental data, with the minimum number of 
parameters and for all $N_{p8}$ bins, was achieved by the use of a simplified, \emph{NKG-like}, LDF 
defined as:
%
\begin{equation} 
  \label{eq:nkglikeeq}
  \rho(r)= A \Big( \frac{r}{r_{0}}\Big)^{s'-2} \Big( 1 + \frac{r}{r_{0}}\Big)^{s'-4.5}
\end{equation}
\vskip -0.10cm
\noindent 
where $A$ is a normalization factor, $s'$ is the \emph{shape parameter} 
that plays the role of the \emph{lateral age},
and $r_0$ is a constant \emph{scale radius}.
 Both simulated and real data suggested the value $r_0 = 30$m as that giving the best $\chi^{2}$ values for the fit
by the previous function in the different $N_{p8}$ bins.
 As a check, different values of $r_0$ were
also used in Eq.\ref{eq:nkglikeeq} when fitting the particle distributions both from data and MC samples.
 For instance, a higher value of $r_0$, like $r_0 = 50$m, also reproduces in a satisfactory way the density lateral
profiles, although with systematically lower values of the parameter $s'$ with respect to the $r_0 = 30$m case. This 
in fact is expected if $r_0$ is actually correlated to the Moli\`ere radius, since higher altitudes (i.e. greater 
$r_0$ values) imply the observation of younger showers (i.e. smaller $s'$ values). This indeed was observed both 
for data and MC events.

 Some of the results obtained using the LDF in Eq.\,\ref{eq:nkglikeeq} (with $r_0 = 30$m) are outlined 
for simulated data (from p, He and Fe initiated showers), in a given $N_{p8}$ bin, 
in Fig.\,\ref{fig:fit_various_func} and Fig.\,\ref{fig:quality_fit_p}, 
where the fits to the average lateral density profiles and their fractional residuals are shown, respectively.
 The fits with the different LDFs given in Eq.\,\ref{eq:nkg_kasc}, Eq.\,\ref{eq:linsley} and Eq.\,\ref{eq:lagutin}, 
together with the related fractional residuals, are also shown for comparison 
(the results from the fit with the LDF given in Eq.\,\ref{eq:nkg} are not shown for the sake of clarity).

  The plot with residuals is also reported in Fig.\,\ref{fig:quality_data} for experimental data, in the same 
$N_{p8}$ bin ($10^{4.7} - 10^{5.0}$) and angular range ($\theta < 15^{\circ}$) already used for MC samples.
 Similar results are obtained in the other $N_{p8}$ bins.
As can be seen, the use of the LDF parametrization as given in Eq.\,\ref{eq:nkglikeeq}
is expected to give the best result for experimental data too.

 Just to give a specific example, 
 the average lateral density distribution for real events in the interval  
$10^{4.7} < N_{p8} < 10^{5.0}$ is reported in Fig.\,\ref{fig:nkg_data} and compared with the corresponding MC 
distributions, from a p-primary event sample (typical proton energy $E_{p} \sim$\,500\,TeV) and from 
an iron-primary event sample ($E_{Fe} \sim$1.4\,PeV).
 As clearly shown, the lateral particle density profile from data is properly fitted by 
the LDF of Eq. \ref{eq:nkglikeeq}, giving a slope parameter value $s' = 1.35$, and lies between the two average 
distributions from pure proton and iron induced showers. The fits to the lateral distributions from simulated
data, by the same \emph{NKG-like} function in Eq.\,\ref{eq:nkglikeeq}, 
give respectively: $s' = 1.17$ for the proton sample and $s' = 1.44$ for the iron sample.
 This indeed is what would be expected, provided that $s'$ reflects the developing stage of the shower, being a detected 
p-induced shower on average younger (which implies a smaller $s'$ value) than a shower induced by an iron nucleus and generating a 
detected event in the same $N_{p8}$ interval.
 In their turn, the real data events in the same $N_{p8}$ interval certainly are produced by a
mixture of primary species, thus giving an intermediate $s'$ parameter value at the detector level.

 An exhaustive study showed that the function in Eq.\,\ref{eq:nkglikeeq} is able to properly fit the lateral particle 
distributions, both from experimental and simulated data, in every $N_{p8}$ bin, up to $\sim 10 \,m$, so
this value of maximum distance from the core was set for the subsequent single event particle
distribution fits.

 By fitting the lateral particle distribution event by event with the function in Eq.\,\ref{eq:nkglikeeq},
a distribution of the $s'$ parameter is obtained which reflects the dependence on both the energy and the mass of the primary particles. 
In Fig.\,\ref{s_distr_G4}, the $s'$ distributions from the fit of single events in proton, helium, CNO
group and iron samples, in the truncated size interval $\Delta N_{p8}=10^{3.4}-10^{3.7}$, are reported.
 In a similar way, the $s'$ distributions from the fit of single events in MC samples of different primaries 
are reportated in Fig.\,\ref{s_distr_G1}, for the truncated size interval of the G1 gain scale 
$\Delta N_{p8}=10^{4.6}-10^{4.9}$.
 The results from the mixed MC samples, obtained using the elemental spectra and composition of the H\"orandel 
\cite{hoerandel2003} and GST \cite{GST2013} models, respectively, are also shown in the two plots.
 In particular, the H\"orandel model is used in weighting individual element distributions in both figures.

In the same plots, the $s'$ distribution obtained from the 
fit of single events in ARGO-YBJ data-sets, from G4 and G1 gain scales, respectively, 
in the previous $N_{p8}$ intervals, are also superimposed for comparison.
 The two truncated size $N_{p8}$ bins have been chosen in such a way that 
the first is included in the G4 gain scale range and the other one in the G1 scale range.

 A discrepancy can be observed in those distributions between MC and real data, which anyway seems to
reduce when the 
$N_{p8}$ values increase and if the GST model is used in the simulated sample instead of the H\"orandel one.
This discrepancy could arise from the elemental composition and spectra used in the simulation models, which
may not reflect the real ones.
 The lowest $N_{p8}$ bin ($log_{10}(N_{p8})=$ [3.4-3.7]) roughly corresponds to $\sim$\,50\,TeV protons and 
$\sim$\,300\,TeV iron nuclei. At these energies the uncertainties on the measured fluxes and spectral indexes
are indeed very large,
specially for the heavy components. 
 All things considered, the disagreement appears smaller in the case of the comparison with the GST model, which indeed
contains larger quantities of heavier elements.
Another possible contribution to the discrepancy could come from the hadronic interaction model
adopted in the simulation. In order to 
evaluate this possible effect, a simulated sample of proton and Helium initiated showers produced by the 
{\it SIBYLL-2.1} \cite{engel1999} interaction model implemented in the CORSIKA code was also used. 
 The comparison of the $s'$ distributions obtained from the {\it QGSJET} and the 
{\it SIBYLL} samples, 
both for G4 and G1 gain scale selection, shows a difference within few percent between the two models.
 On the other hand, Figures \ref{s_distr_G4} and \ref{s_distr_G1} are related to the very forward kinematic region of the shower
development. 
 Thus part of the disagreement could be due to a not perfect description of the hadronic interaction in
that region.

 It is important here to underline that the above discussed disagreement does not compromise at all 
the results of the analysis which is carried out and discussed in the following sections.
  
An accurate study was also performed on the possible dependence of the age parameter $s'$ on the
radial distance from the core. A slight variation was found in the considered range of distances up
to 10 m at maximum, thus allowing in a reliable way to take the parameter value from the fit 
in the whole range as a sort of average local age.

\section{Shower age determination}
\label{sec:universal}

 \vskip -0.3cm
 From Fig.\,\ref{s_distr_G4} and Fig.\,\ref{s_distr_G1} it can be seen that, for a fixed interval of $N_{p8}$, 
the $s'$ distribution gradually moves towards higher values as far as the primary mass
increases, namely going from proton to iron, as a consequence of a larger primary interaction cross section with
atmosphere nuclei.
 Conversely, for a given primary, the fit parameter $s'$ 
values decrease when $N_{p8}$ (i.e. the energy) increases, thus suggesting the observation 
of deeper showers at larger energies. Furthermore, as expected \cite{bib:capdev2012}, 
the proton distribution appears wider than those of heavier primaries.

\begin{figure}
  \centerline{\includegraphics[width=0.7\textwidth, height=6.5cm]{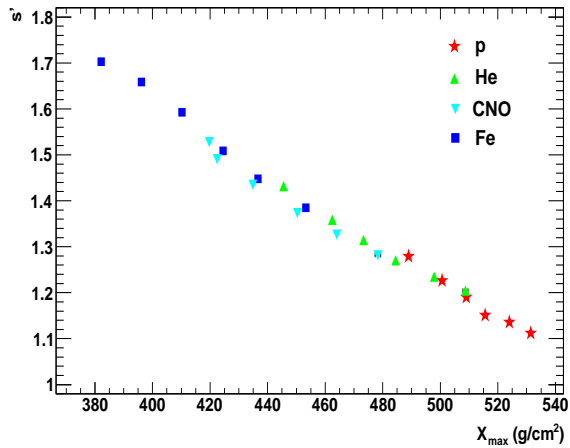} }
 \caption{The average lateral age parameter $s'$ resulting from the fits of the lateral particle
          distributions in single events of simulated p, He, CNO group and Fe samples (in each $N_{p8}$ bin, see text) 
	  vs the corresponding $X_{max}$ average values. Only near-vertical showers 
	  ($\theta < 15^{\circ}$) are considered.} 
 \label{s_Xmax_pHeCNOFe}
\end{figure}
\begin{figure}
  \centerline{\includegraphics[width=0.7\textwidth, height=6.5cm]{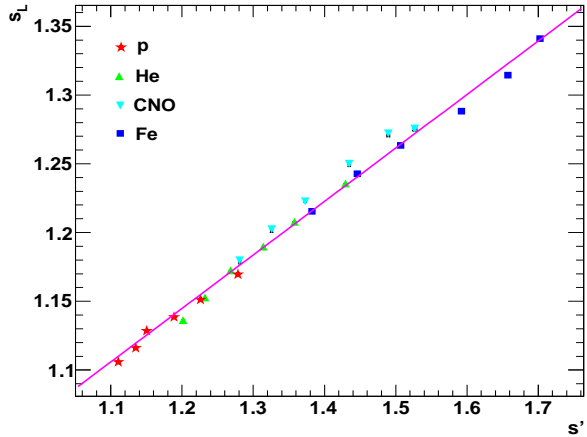} }
 \caption{Average longitudinal age $s_{L}$ vs the corresponding lateral age parameter $s'$ 
          resulting from the fits of the single event lateral particle distributions for simulated p, He, CNO group 
	  and Fe samples (in each $N_{p8}$ bin, see text), in the zenith angle range $\theta < 15^{\circ}$.}
 \label{s_sL_pHeCNOFe}
\end{figure}

 Such features are in agreement with the expectations, the slope $s'$ being correlated with 
the shower age, thus reflecting its development stage.
 In other words, in this context the $s'$ parameter plays the role of {\it lateral age}.
 As a consequence, if we plot the fit $s'$ values, for each simulated primary type and several $log_{10}(N_{p8})$ 
intervals (namely: $log_{10}(N_{p8})=$ [3.4-3.7], [3.7-4.0], [4.0-4.3], [4.3-4.6], [4.6-4.9], [4.9-5.2]), as a 
function of the corresponding $X_{max}$ average values, we obtain the
correlation shown in Fig.\,\ref{s_Xmax_pHeCNOFe}. 
 The $X_{max}$ value for each event is that provided by CORSIKA as a result of the shower longitudinal
profile fit by means of the Gaisser-Hillas function \cite{gais_hill}.
 Each point in the plot represents a distribution, whose uncertainty on the mean (which is very small) has been
considered for the graph, although large event by event fluctuations occur, so the RMSs of the distributions
are quite big. Anyway, here we are only interested to the behaviour of the average correlation.
As can be seen, the shape parameter $s'$ depends only on the development stage of the shower, 
independently from the nature of the primary particle and energy.
That plot expresses an important 'universality property' of the 
detected shower development in the atmosphere, in terms of the age parameter given by the LDF slope.
 This also implies the possibility to select the most deeply penetrating showers at different zenith angles, an
important point for correlating the exponential angular rate distribution with the interaction length of the 
impinging particle \cite{cross_sec}.

\begin{figure}[t]
  \centerline{\includegraphics[width=0.7\textwidth, height=6.5cm]{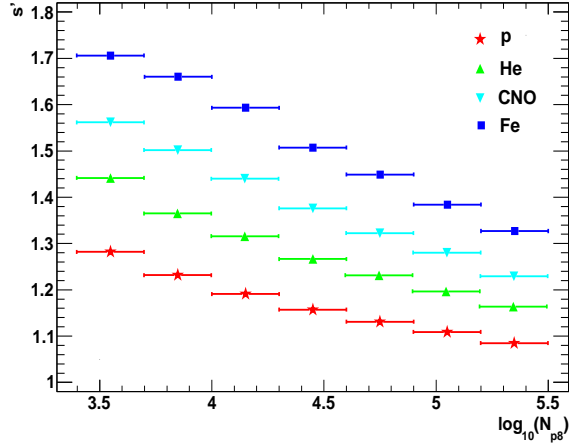} }
 \caption{Average lateral age parameter s' as resulting from the fits of lateral particle distributions of simulated p,
 He, CNO group and Fe induced showers, in several $N_{p8}$ bins, using the infunction in Eq.\,\ref{eq:nkglikeeq}.
 The events are selected with the same cuts of real ARGO-YBJ data. The error on the mean is considered for each point 
 in the plot.}
 \label{s_Np8_pHeCNOFe}
\end{figure}
\begin{figure}[t!]
  \centerline{\includegraphics[width=0.7\textwidth, height=6.5cm]{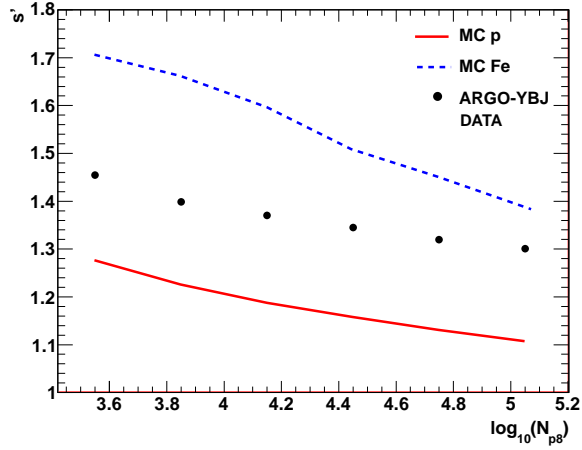} }
 \caption{Average lateral age parameter $s'$ in several $N_{p8}$ bins, as resulting from the fits of lateral particle 
 distributions of ARGO-YBJ data events (black dots). 
 The $s'$ behaviours for simulated p (red solid line) and iron (blue dashed line) induced showers 
 relying on the H\"orandel model 
 are also reported.
 The error on the mean is considered for each point, while the related $s'$ distribution is
 much wider ($RMS \sim 0.2$) due to the large fluctuations.
 (For interpretation of the references to colour in this figure legend, the reader is referred to the 
 web version of this article.)}
 \label{s_Np8_data}
\end{figure}

 Given its nature of lateral age, the parameter $s'$ is also expected to be strongly related to the 
longitudinal age ($s_{L}$) of the shower.
 Such correlation is made more explicit using the classical definition of $s_{L}$ as a function of 
the shower maximum depth $X_{max}$:
  
\begin{equation} 
  \label{eq:s_long}
  s_{L} = \frac{3 h_0 \cdot sec(\theta)}{h_0 \cdot sec(\theta) + 2 X_{max}}
\end{equation}

\noindent 
where $h_0$ is the detector vertical atmospheric depth, $\theta$ the shower zenith angle. 
Then, using the same average $X_{max}$ values reported in Fig.\,\ref{s_Xmax_pHeCNOFe}, through the 
previous relation we obtain the plot of Fig.\,\ref{s_sL_pHeCNOFe}, which clearly demonstrates 
how the 'observed' lateral age $s'$ is strictly related to the longitudinal age $s_{L}$, and moreover a linear 
dependence appears appropriate.
 Indeed, in the same plot a linear fit has been superimposed, which gives:
\begin{equation} 
  \label{eq:sl_fit}
  s_{L} = (0.389 \pm 0.005) \cdot s' + (0.678 \pm 0.007)
\end{equation}

 To notice that such relation between $s'$ and $s_{L}$, clearly stated on the averages, is also valid on the single
event basis, apart small fluctuations mainly introduced by the fit uncertainties (it was found that $s_{L}$ can be 
obtained event-by-event by Eq.\,\ref{eq:sl_fit} with a resolution of about 6$\,\%$ independently of the mass). 

  We can more deeply investigate the above described results and get further consequences:
the universality property expressed by the plot in Fig.\,\ref{s_Xmax_pHeCNOFe} allows to decouple the 
detected shower signal from the primary nature thanks to the (linear) relation $s'=s'(X_{max})$, once the 
$s'$ parameter value was obtained from the single event LDF fit. 
 This in fact suggests the possibility to exploit such $s'$ properties in order to identify a mass independent primary
energy estimator.

\section{Sensitivity to primary mass}
\label{sec:composition}
 In the previous sections, the main features of the shower age parameter $s'$ were discussed, sufficient to reveal its
sensitivity to the mass of primary particles.
More explicitely, from the fit of the lateral particle distribution of single events in the simulated samples of each 
primary (p, He, CNO, Fe), it was found that the age parameter $s'$ value decreases when $N_{p8}$ (i.e. the energy) 
increases, this being due to the observation of {\it younger} (deeper) showers at larger energies.
In the meantime, for a given range of $N_{p8}$, the average $s'$ increases going from hydrogen to iron, as a 
consequence of a larger primary interaction cross section with atmosphere nuclei
producing showers which on average have a flatter lateral profile at the detection level.
This is summarized in Fig.\,\ref{s_Np8_pHeCNOFe}, which shows the average $s'$ values obtained for the whole simulated
samples of hydrogen, helium, CNO group and iron nuclei.
 
 The straightforward implication of this is that $s'$ from the LDF fit very close to the shower 
axis, together with the measurement of the truncated size $N_{p8}$, can give information on 
the nature of the particle initiating the cascade,
thus making possible the study of CR primary mass composition.

 From the LDF fits of real data events with the function in Eq.\,\ref{eq:nkglikeeq}, in the same $N_{p8}$ intervals used
for MC data, similar $s'$ distributions are obtained. 
The average $s'$ values from ARGO-YBJ experimental data are reported in Fig.\,\ref{s_Np8_data}, together with the 
corresponding fit results from MC simulations for protons and iron initiated showers (''extreme pure
compositions'').
Each point of course is the mean of a distribution, whose width is quite large 
(the $RMS$ varies between $0.16$ and $0.25$) 
due to the shower by shower fluctuations,
while the error on the mean is very small (such errors are associated to the dots in the plot).
The experimental data points nicely lie between the expectations from extreme pure compositions, roughly indicating a
mixed composition becoming gradually heavier when the primary energy increases.

\section{Conclusions}
\label{sec:conclu}
 A detailed study of the lateral particle distribution around the axis of the quasi-vertical extensive air showers 
detected by the ARGO-YBJ experiment has been performed. The analysis of data triggering the RPC charge readout system allowed to
explore a wide range of particle density, from few particles /m$^2$ up to several $10^4$/m$^2$ very near to the core.
 A \emph{NKG-like} function has been identified as LDF, which is able to properly describe the particle distribution up to about 
10$\,$m from the core, both for simulated and experimental data. 
 Applied to simulated data originated by different primaries, this study showed how the slope of such LDF,
given by a {\it lateral age} fit parameter $s'$, gives information on the longitudinal shower development.
 Moreover, it demonstrated the existence of an important
universality property of the shower development stage features, when expressed in terms of $s'$, with respect to the primary nature.
 Thus, the fit by a proper LDF to the lateral density profile of the events as detected by an array like 
ARGO-YBJ, although limited to $\sim 10 \,m$ around the axis, provides an effective tool to determine the shower age.
 
 The sensitivity of the particle distribution shape, as measured by ARGO-YBJ within few meters around the core, to the nature 
of the primary particle generating the shower, has also been demonstrated and discussed.
 This suggests the possibility of using the lateral age parameter $s'$, the slope of the LDF which describes such lateral density 
profile, for the study of the CR mass composition. In particular, that parameter could provide a new and efficient way
of selecting samples of events initiated by light mass primaries (i.e. protons and alpha
particles), without relying on the muon signal, thus avoiding sizeable systematic dependencies on
the adopted hadronic interaction model.

 It is worthwhile to emphasize once again that the results presented and discussed in this paper
have been achieved thank to the peculiar layout of the ARGO-YBJ detector as a 'full-coverage carpet', as well as
to the implementation of the RPC analog charge readout system, which allowed to 
measure the particle density distribution very close to the detected shower core.

\vskip 0.5cm

\centerline{\bf Acknowledgments}
This work is supported in Italy by the Istituto Nazionale di Fisica Nucleare (INFN) and the Ministero dell'Istruzione, 
dell'Universit\`a e della Ricerca (MIUR) and in China by NSFC (No. 10120130794 and No. 11575203), the Chinese Ministry of Science and Technology, 
the Chinese Academy of Sciences, the Key Laboratory of Particle Astrophysics, CAS.
 We also aknowledge the
essential support of W.Y. Chen, G. Yang, X.F. Yuan, C.Y. Zhao, R. Assiro, B. Biondo, S. Bricola, F. Budano, A.
Corvaglia, B. D'Aquino, R. Esposito, A. Innocente, A. Mangano, E. Pastori, C. Pinto, E. Reali, F. Taurino and A.
Zerbini, in the installation, debugging and maintenance of the detector.

 

\end{document}